\documentclass[a4paper]{jpconf}
\pdfoutput=1

\usepackage{cite} 
\usepackage{graphicx}
\usepackage{bm}
\usepackage{amssymb,amsmath,latexsym}
\usepackage{color}
\definecolor{darkblue}{RGB}{0,0,196}
\definecolor{darkred}{RGB}{196,0,0}
\usepackage[colorlinks=true,linktocpage=true,linkcolor=darkblue,citecolor=darkblue,urlcolor=darkblue]{hyperref}

\newcommand{\p}{\mathbf{p}}

\def\be{\begin{equation}}
\def\ee{\end{equation}}
\def\ba{\begin{eqnarray}}
\def\ea{\end{eqnarray}}

\newcommand*\barM{{\overline{\cal M}\hspace{0.5mm}}} 

\begin{document}

\title{Pseudothermalization of the quark-gluon plasma}

\author{Michael Strickland}

\address{Department of Physics, Kent State University, Kent, OH 44242, United States}
\ead{mstrick6@kent.edu}

\begin{abstract}
In this proceedings contribution I review recent work in kinetic theory which demonstrates that, for system undergoing Bjorken expansion, there exists an attractor in all moments of the one-particle distribution function.  I discuss how this attractor emerges in both exact solutions obtained in relaxation time approximation (RTA) and the effective kinetic theory approach to high-temperature quantum chromodynamics (QCD).  The QCD effective kinetic theory collisional kernel used includes both elastic ($2 \leftrightarrow 2$) and LPM-resummed inelastic ($2 \leftrightarrow 1$) contributions.  The results obtained indicate that a pseudothermal attractor exists in both RTA and QCD kinetic theory and that their respective attractors can be extended to early times when the system is far from equilibrium.  Finally, I discuss how knowledge of the QCD effective kinetic theory attractor can be used to assess different hydrodynamic freeze-out prescriptions used in heavy-ion phenomenology.  The results obtained show that improved freeze-out prescriptions such as anisotropic hydrodynamics perform better in conditions corresponding to those generated in high-multiplicity pA and pp collisions, e.g. short lifetime and high inverse Reynolds number.
\end{abstract}

\section{Introduction}

\vspace{3mm}

In recent years there has been intense interest in the degree to which the dynamics of non-equilibrium quantum field theories can be reliably described by relativistic viscous hydrodynamics.  This has impacts on our understanding on the quark-gluon plasma (QGP) generated in relativistic heavy-ion collisions \cite{Averbeck:2015jja,Jeon:2016uym,Romatschke:2017ejr}, condensed matter systems in which a hydrodynamical description appears to be successful \cite{Sachdev:2010ch,2017stmc.bookN}, and potentially studies of neutron star mergers \cite{Duez:2004nf,Shibata:2017xht,Most:2018eaw,Alford:2019kdw}.  In the context of heavy-ion collisions (AA) the fundamental question is how well can quarks and gluons produced in the first fractions of a fm/c be described by relativistic hydrodynamics and whether or not such a description can be extended to collisions of small systems such as pA and pp.  In the context of AA collisions first indications for the effectiveness of relativistic viscous hydrodynamics came from the AdS/CFT studies of Chesler and Yaffe \cite{Chesler:2009cy} wherein it was shown that, despite rather large early-time pressure anisotropies in the local rest frame of the system, the dynamics of the energy-momentum tensor could be well-described by viscous hydrodynamics.  This implied that the system does not rapidly thermalize and instead rapidly {\em hydrodynamizes}.

Since the work of Chesler and Yaffe there have been many papers addressing the phenomenon of hydrodynamization which have found that, in a variety of contexts, there exists a ``non-equilibrium attractor'' for the components of the energy-momentum tensor which can be well-described by relativistic dissipative hydrodynamics after a short amount of time in the center of the fireball ($\lesssim 1$ fm/c) \cite{Heller:2015dha,Keegan:2015avk,Heller:2016rtz,Florkowski:2017olj,Romatschke:2017vte,Bemfica:2017wps,Spalinski:2017mel,Romatschke:2017acs,Behtash:2017wqg,Florkowski:2017jnz,Florkowski:2017ovw,Strickland:2017kux,Almaalol:2018ynx,Denicol:2018pak,Behtash:2018moe,Strickland:2018ayk,Heller:2018qvh,Behtash:2019qtk,Strickland:2019hff,Jaiswal:2019cju,Kurkela:2019set,Chattopadhyay:2019jqj,Brewer:2019oha}. In the majority of these studies, the authors concentrated on the evolution of the energy-momentum tensor, which in a kinetic theory language translates into a particular set of moments of the one-particle distribution function.  A natural question which arises is whether or not a non-equilibrium attractor can be observed in higher moments of the distribution function or, even more generally, the distribution function itself.  The first attempts to address this question was presented in Refs.~\cite{Strickland:2018ayk,Strickland:2019hff}, where it was shown that exact solutions to relaxation time approximation (RTA) Boltzmann equation in a Bjorken-expanding background exhibited an attractor in all computed moments of the distribution function.  

In this proceedings contribution, I review past work using exact solutions to the RTA kinetic theory in systems undergoing Bjorken expansion.  I then present similar findings reported recently in Ref.~\cite{Almaalol:2020rnu} using the effective kinetic theory (EKT) approach to the dynamics of a non-equilibrium quark-gluon plasma \cite{Arnold:2002zm,Kurkela:2015qoa,Keegan:2015avk}. The EKT evolution used includes both elastic and inelastic contributions to the collisional kernel allowing for a numerical realization of the bottom-up thermalization scenario \cite{Baier:2000sb}.

\section{Exact RTA attractor and dynamics}

\vspace{3mm}

For a system subject to boost-invariant and transversally homogenous Bjorken flow and subject to a RTA collisional kernel,  the underlying kinetic equation is simple
\be
p^\mu \partial_\mu  f(x,p) =  \frac{p \cdot u}{\tau_{\rm eq}(\tau)} \left( f_{\rm eq}-f \right) , 
\label{kineq}
\ee
where $\tau_{\rm eq}(\tau) = 5\bar\eta(\tau)/T(\tau)$ is the relaxation time with $\bar\eta(\tau)=\eta(\tau)/s(\tau)$ being the shear viscosity to entropy density ratio and $T(\tau)$ being the local effective temperature.  Eq.~\eqref{kineq} can be cast into simpler form by writing it in terms of manifestly boost-invariant variables~\cite{Bialas:1984wv,Bialas:1987en}.  The resulting simpler equation can easily be shown to have a general solution given by~\cite{Baym:1984np,Florkowski:2013lza,Florkowski:2013lya}
\be
f(\tau,w,p_T) = D(\tau,\tau_0) f_0(w,p_T) 
+  \int_{\tau_0}^\tau \frac{d\tau^\prime}{\tau_{\rm eq}(\tau^\prime)} \, D(\tau,\tau^\prime) \, 
f_{\rm eq}(\tau^\prime,w,p_T) \, ,  
\label{eq:exactsolf}
\ee
where $w = t p_L - z E$ and $D$ is the damping function
\be
D(\tau_2,\tau_1) = \exp\left[-\int\limits_{\tau_1}^{\tau_2}
\frac{d\tau^{\prime\prime}}{\tau_{\rm eq}(\tau^{\prime\prime})} \right] .
\ee

Equation~\eqref{eq:exactsolf} can be turned into an infinite tower of equations for moments of the one-particle distribution function
\be
{\cal M}^{nm}[f] \equiv \int dP \,(p \cdot u)^n \, (p \cdot z)^{2m} \, f(\tau,w,p_T) \, .
\label{eq:genmom1}
\ee
Assuming classical statistics, one obtains~\cite{Strickland:2018ayk}
\ba
{\cal M}^{nm}(\tau) &=& \frac{\Gamma(n+2m+2)}{(2\pi)^2} \Bigg[ D(\tau,\tau_0) T_0^{n+2m+2} \frac{{\cal H}^{nm}\!\left( \tfrac{\alpha_0 \tau_0}{\tau} \right)}{[{\cal H}^{20}(\alpha_0)/2]^{(n+2m+2)/4}} \nonumber \\
&& \hspace{1cm} 
+ \int_{\tau_0}^\tau \frac{d\tau^\prime}{\tau_{\rm eq}(\tau^\prime)} \, D(\tau,\tau^\prime) \, 
T^{n+2m+2}(\tau') {\cal H}^{nm} \hspace{-1mm} \left( \tfrac{\tau'}{\tau} \right) \Bigg] ,
\label{eq:meqfinal}
\ea
with
\be
{\cal H}^{nm}(y) = \tfrac{2y^{2m+1}}{2m+1}  {}_2F_1(\tfrac{1}{2}+m,\tfrac{1-n}{2};\tfrac{3}{2}+m;1-y^2)  \, .
\ee
Note that certain moments map to familiar hydrodynamics variables, e.g. $n = {\cal M}^{10}$, $\varepsilon = {\cal M}^{20}$, and $P_L = {\cal M}^{01}$.

One can obtain a closed integral equation for $T(\tau)$ by considering the integral equation obeyed by ${\cal M}^{20} = \varepsilon = \varepsilon_{\rm eq}$ which simplifies to
\be
T^4(\tau) = D(\tau,\tau_0) T_0^4 \frac{{\cal H}\!\left( \frac{\alpha_0 \tau_0}{\tau} \right)}{{\cal H}(\alpha_0)} \nonumber \\
+ \int_{\tau_0}^\tau \frac{d\tau^\prime}{2 \tau_{\rm eq}(\tau^\prime)} \, D(\tau,\tau^\prime) \, 
T^4(\tau') {\cal H}\hspace{-1mm} \left( \frac{\tau'}{\tau} \right) . 
\label{t4eq}
\ee
This equation can be numerically solved iteratively~\cite{Florkowski:2013lza,Florkowski:2013lya}.  Once the solution for $T(\tau)$ is obtained, one can use this to solve for all other moments ${\cal M}^{nm}(\tau)$ using Eq.~\eqref{eq:meqfinal} and the full distribution function itself using Eq.~\eqref{eq:exactsolf}.

\begin{figure}[t!]
\includegraphics[width=1\linewidth]{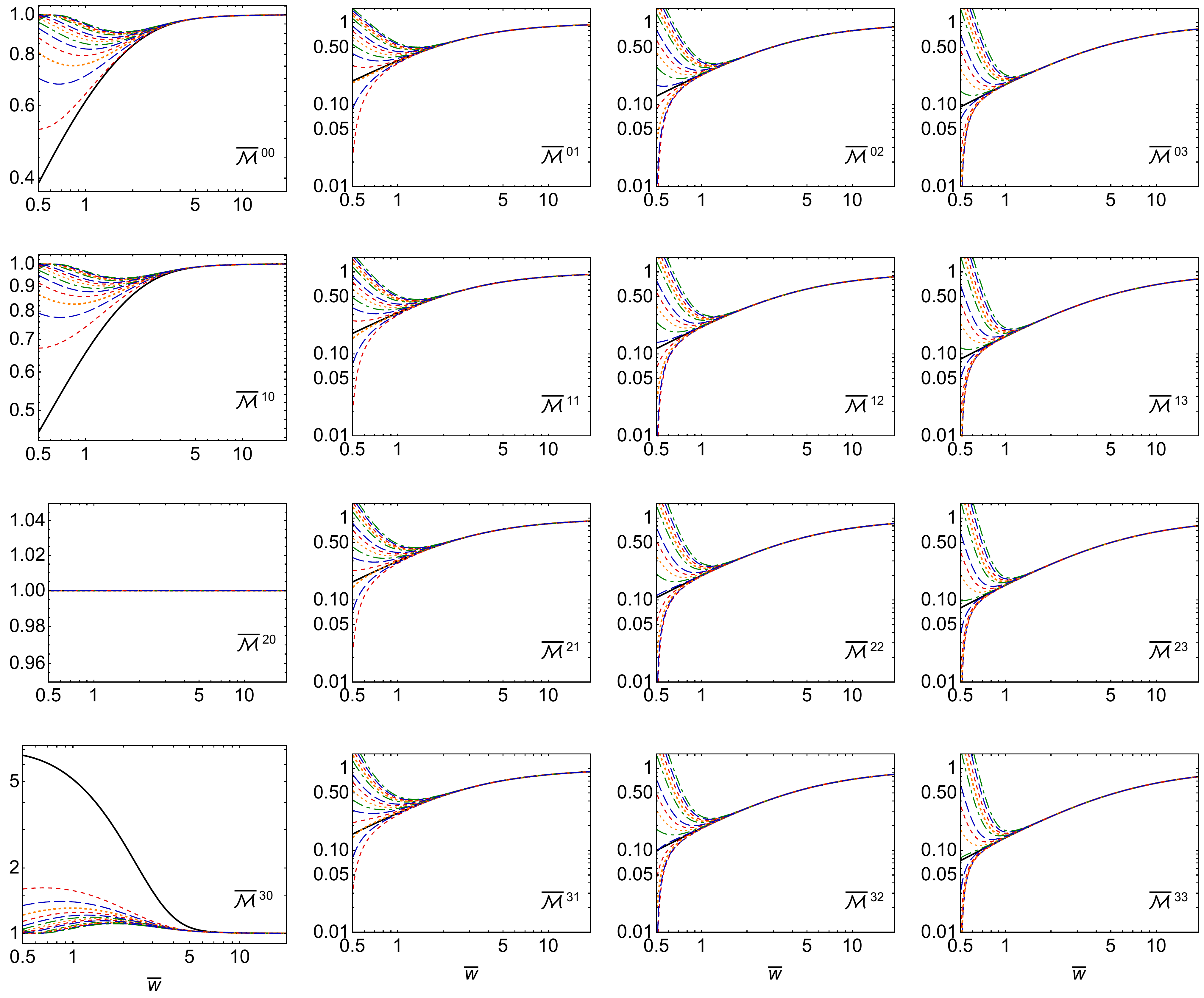}
\caption{Scaled moments $\barM^{nm}$ obtained from the exact RTA attractor solution (solid black line) compared to a set of exact RTA solutions (various colored dotted and dashed lines) initialized at $\tau = 0.1$ fm/c with varying initial pressure anisotropy.  The horizontal axis is $\overline{w} \equiv \tau/\tau_{\rm eq} = \tau T/5 \bar\eta$.  Panels show a grid in $n$ and $m$.}
\label{attractorGridSols}
\end{figure}

\begin{figure}[t!]
\includegraphics[width=1\linewidth]{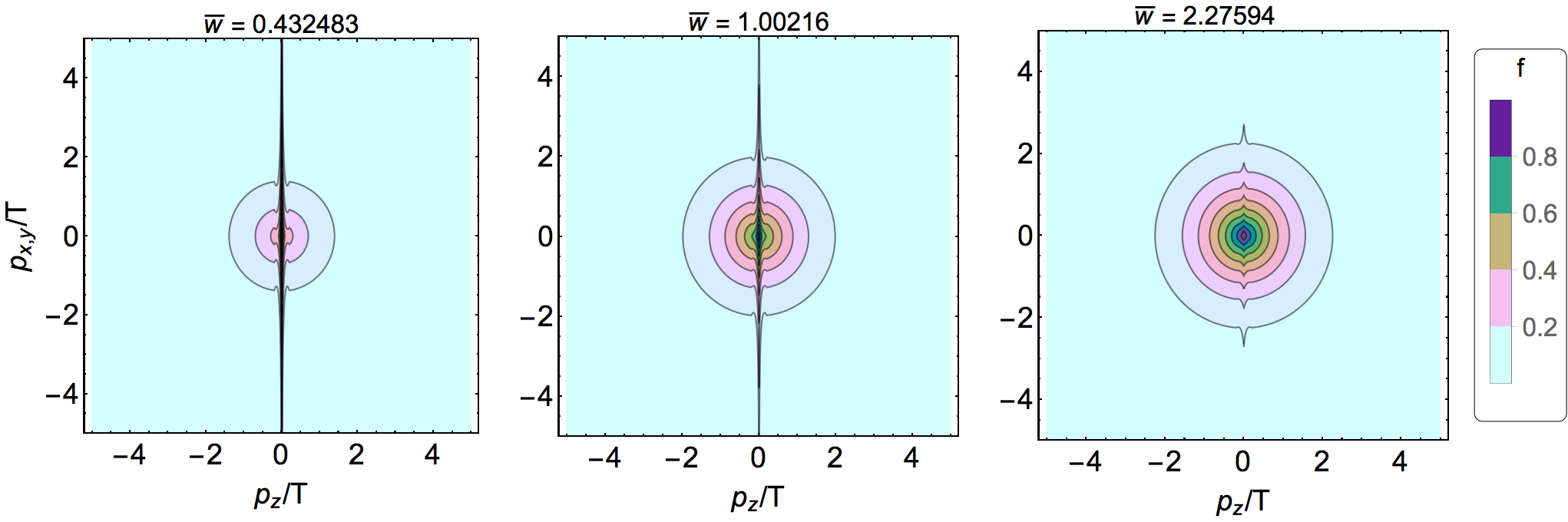}\\
\includegraphics[width=1\linewidth]{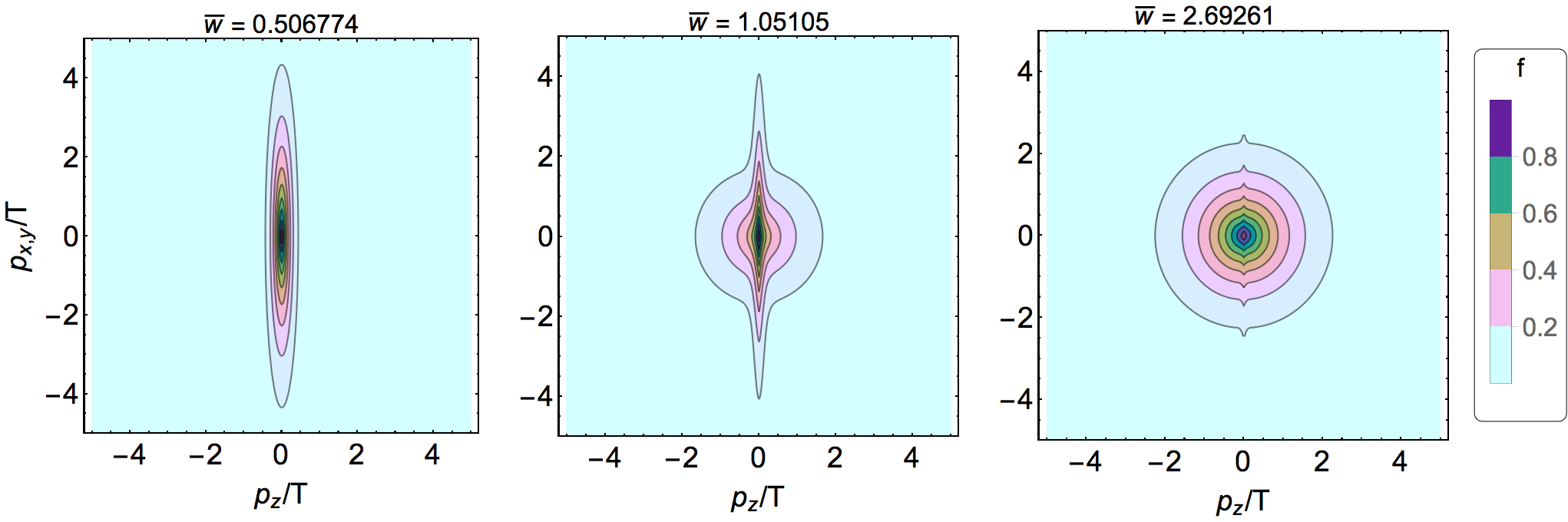}
\caption{(Top) Visualization of the one-particle distribution function associated with the RTA attractor. (Bottom) Visualization of the one-particle distribution function obtained using a typical (non-attractor) anisotropic initial condition and an RTA collisional kernel.}
\label{f_vis1}
\end{figure}

In Fig.~\ref{attractorGridSols} I present the results obtained originally in Ref.~\cite{Strickland:2018ayk}.  Each panel shows the dependence of a scaled moment of the distributions function, $\overline{\cal M}^{nm} \equiv {\cal M}^{nm}/{\cal M}^{nm}_{\rm eq}$.  The black line in each panel is the exact attractor for that moment of the distribution function and the colored dashed lines are solutions with particular momentum-space anisotropic initial conditions.  As can be seen from these panels, for all moments with $n>1$ the solutions converge to the attractor solution by approximately $\overline{w} = \tau/\tau_{\rm eq} = 2$.  Closer analysis reveals that higher order moments converge to their respective attractors more quickly \cite{Strickland:2018ayk,Strickland:2019hff}.  The convergence of moments with $n=0$ is slower and can be traced to the free streaming contribution to the solution \cite{Strickland:2018ayk,Strickland:2019hff}.  Overall the results indicate that there is an attractor for the entire distribution function.

In Fig.~\ref{f_vis1} I present contour plots of the corresponding one-particle distribution functions obtained using Eq.~\eqref{eq:exactsolf}.  The top panels show the distribution function obtained along the exact RTA attractor and the bottom panels show a typical evolution obtained with a spheroidal initial condition.  The columns from left to right correspond to different snapshots in rescaled time $\overline{w}$.  As can be seen from this Figure, the bottom panels approach the form given by the attractor solution presented in the top panels, converging to one another at rather short rescaled time.  For more details concerning the nature of the attractor for the full one-particle distribution function see Ref.~\cite{Strickland:2018ayk}.

\section{Effective kinetic theory approach to QCD}

\vspace{3mm}

RTA is a nice toy model in which it is possible to obtain the exact solutions presented above, however, it would be better to ask the same questions in QCD.  In Ref.~\cite{Almaalol:2020rnu} we made  use of a numerical implementation of the Arnold, Moore, and Yaffe effective kinetic theory (EKT) \cite{Arnold:2002zm} which allows for the description of both thermally occupied and over-occupied gluonic plasmas \cite{York:2014wja,Kurkela:2015qoa}.  In practice, for a system undergoing boost-invariant Bjorken expansion, the code solves an EKT Boltzmann equation of the form
\be
-\frac{d f(\p)}{d\tau} = \mathcal{C}_{1\leftrightarrow 2}[f(\p)] + \mathcal{C}_{2\leftrightarrow 2}[f(\p)] + \mathcal{C}_{\rm exp }[f(\p)] \, ,
\label{eq:ekt1}
\ee
where $f(\p)$ is the gluonic one-particle distribution function.  We discretize $\p$ in spherical coordinates, $p = |{\bf p}|$, \mbox{$x \equiv \cos\theta$}, and $\phi$.  The effect of longitudinal expansion is included through  $\mathcal{C}_{\rm exp}[f(\p)] = - \frac{p_z}{\tau}\frac{\partial }{\partial p_z} f(\p)$ \cite{Mueller:1999pi}.  The elastic scattering term $\mathcal{C}_{2\leftrightarrow 2}$ and the effective inelastic term $\mathcal{C}_{1 \leftrightarrow 2}$ include physics of dynamical screening and Landau-Pomeranchuck-Migdal (LPM) suppression and, in order to find the form of the collision kernels, self-energy and ladder resummations are required. For details,  see Refs.~\cite{Arnold:2002zm,York:2014wja,Kurkela:2015qoa}. 

\begin{figure}[t!]
\includegraphics[width=1\linewidth]{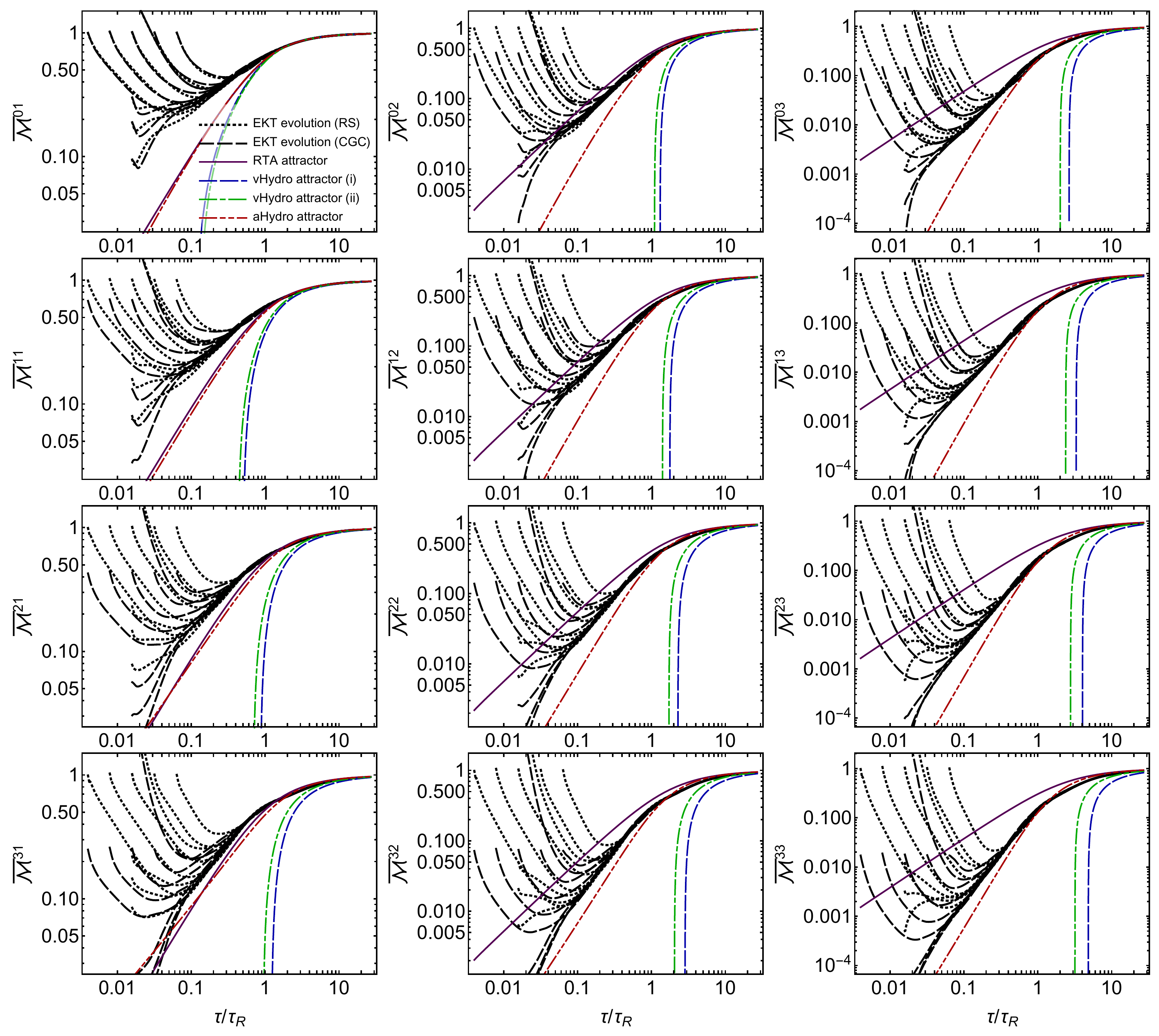}
\caption{Evolution of the scaled moments $\overline{\cal{M}}^{nm}$ with $n \in \{0,1,2\}$ and $m \in \{0,1,2,3\}$.  Black dotted and dashed lines show EKT evolution with RS amd CGC initial conditions, respectively.  The purple solid line is the exact RTA attractor, the blue long-dashed line is the DNMR vHydro attractor using $\delta f$ parameterization (i), the green dot-dashed line is the DNMR vHydro attractor using $\delta f$ parameterization (ii), and the red dot-dot-dashed line is the aHydro attractor.}
\label{supfig1}
\end{figure}

For the numerical solution of Eq.~\eqref{eq:ekt1}, we discretized $n(\p) = p^2 f(\p)$ on a three-dimensional grid in momentum space and used Monte Carlo sampling to compute the integrals appearing in the elastic and inelastic collisional kernels. The algorithm used is based on Ref.~\cite{York:2014wja} and exactly conserves energy while also exactly accounting for the particle number violation originating from the inelastic contributions to the collisional kernel.  Due to the azimuthal symmetry of Bjorken flow,  we used an effectively two-dimensional grid:  250 $\times$ 2000 points in the $p$ and $x=\cos\theta$ directions, respectively. 

We computed the time evolution of a complete set of integral moments characterizing the momentum dependence of the distribution function \eqref{eq:genmom1}.  As mentioned previously, the energy density is given by $\varepsilon=\nu {\cal M}^{20}$, longitudinal pressure  by $P_L = \nu {\cal M}^{01}$, and number density by $n=\nu {\cal M}^{10}$ for $\nu$ degrees of freedom ($\nu = 2 d_A$ for $d_A$ adjoint colors of gluons).  The other moments do not have an interpretation in terms of the usual hydrodynamic moments considered in the literature.\footnote{The $m=0$ modes are simply related to the effective temperatures introduced in Refs.~\cite{Kurkela:2018xxd,Kurkela:2018oqw}.}  As with the RTA, these moments will be scaled by their corresponding equilibrium values with \mbox{$\overline{\cal{M}}^{nm}(\tau) \equiv {\cal M}^{nm}(\tau) / {\cal M}^{nm}_{\rm eq}(\tau)$}, where, using a Bose distribution, one obtains
\be
{\cal M}^{nm}_{\rm eq} =  \frac{ T^{n+2m+2}\Gamma(n+2m+2) \zeta(n+2m+2)}{2 \pi^2 (2m+1)} \, .
\ee
The temperature $T$ here corresponds to the temperature of an equilibrium system with the same energy density, given by 
$T = (30 \varepsilon/\nu \pi^2)^{1/4}$.  Note that the different moments are sensitive to different momentum regions of the distribution function and for future comparisons, we note that, in equilibrium, the typical momentum contributing to a given moment is  $\langle p \rangle^{nm}_{\rm eq} = {\cal M}^{n+1,m}_{\rm eq}/{\cal M}^{nm}_{\rm eq}$, giving, \emph{e.g.}, \mbox{$\langle p \rangle^{10}_{\rm eq} \simeq 2.7 \, T$},  $\langle p \rangle^{01}_{\rm eq} = \langle p \rangle^{20}_{\rm eq} \simeq 3.83 \, T$, $\langle p \rangle^{21}_{\rm eq} \simeq 5.95 \, T$, and $\langle p \rangle^{33}_{\rm eq} \simeq 11 \, T$.

\begin{figure}[t!]
\includegraphics[width=1\linewidth]{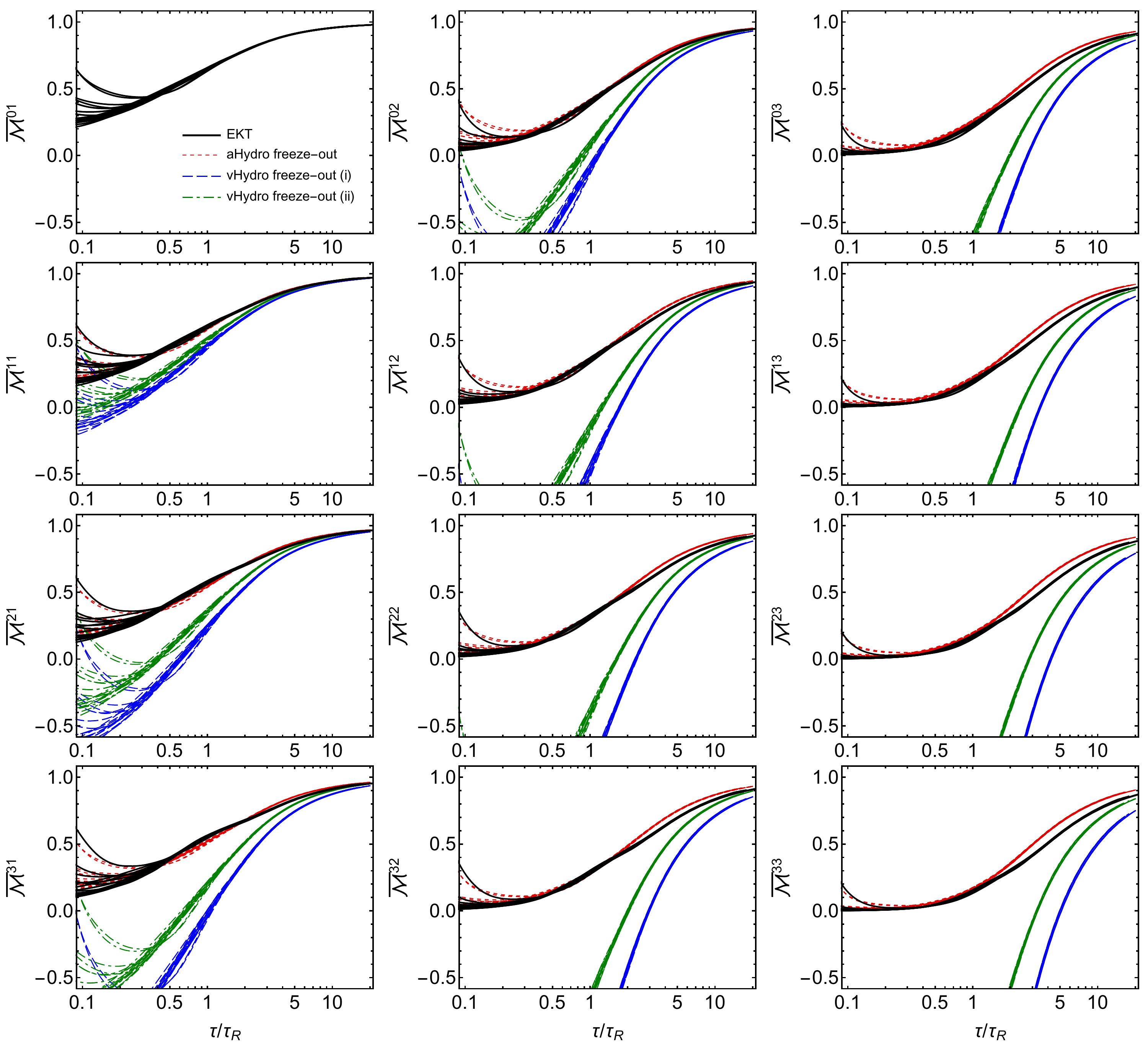}
\caption{Evolution of the scaled moments $\overline{\cal{M}}^{nm}$ with $n \in \{0,1,2\}$ and $m \in \{0,1,2,3\}$ for all runs (CGC +RS) shown as black lines. Other lines -- red-dashed, blue long-dashed, green dot-dashed -- correspond to results obtained in three freeze-out scenarios, aHydro, vHydro (i), and vHydro (ii), respectively. Note that all scenarios agree on $\overline{\cal M}^{01}$ by construction ($P_L$-matching). The other panels are predictions based on this matching.}
\label{supfig2}
\end{figure}

In Fig.~\ref{supfig1} I present a panel of results for the scaled moments.  The horizontal (time) axis is scaled by the instantaneous interaction time of the system $\tau/\tau_R(\tau)$, which measures the age of the system in units of the instantaneous interaction time, $\tau_R(\tau) \equiv 4 \pi \bar\eta/T(\tau)$.  In all panels the black dotted and dashed lines correspond to the EKT results obtained using two different types of initial conditions:  (i) spheroidally deformed thermal (RS) and (ii) over-occupied CGC-type initial conditions.  For details of the initial conditions used see Ref.~\cite{Almaalol:2020rnu}.  For both types of initial conditions, we varied both the initial momentum-space anisotropy and initialization time.  In all panels we also show the attractors corresponding to RTA, viscous hydrodynamics (vHydro) with a quadratic freeze-out form (i), vHydro with a LPM-modified freeze-out form (ii), and anisotropic hydrodynamics \cite{Strickland:2017kux}.

As can be seen from Fig.~\ref{supfig1}, one finds that all EKT solutions afpproach a universal EKT attractor irrespective of the initial conditions assumed.  In particular, we observe that, as the initialization time is decreased, the solutions approach this attractor more rapidly.  This establishes the existence of an early-time or pull-back \cite{Behtash:2019qtk} attractor in EKT QCD.  We additionally observe that, while all approximate attractors (RTA, vHydro, and aHydro) do a reasonable job in reproducing low-order moments, e.g. $P_L/P_L^{\rm eq} = \overline{\cal M}^{01}$, the different schemes do not accurately describe the EKT QCD attractor.  Of the set of approximations, aHydro is the best in reproducing the EKT QCD attractor.

Having determined the EKT QCD attractor, one can use this to assess the ability of different viscous hydrodynamics freeze-out prescriptions in reproducing the behavior observed in the high-order moments.  For this purpose one can determine the shear viscous correction, $\Pi$ or $\xi$ for vHydro and aHydro, respectively, that allows one to exactly reproduce the EKT QCD attractor for $P_L/P_L^{\rm eq} = \overline{\cal M}^{01}$.  Using the extracted $\Pi$ or $\xi$ one can than make predictions for high-order moments.\footnote{For details concerning the freeze-out prescriptions used, see Ref.~\cite{Almaalol:2020rnu}.} I present the results of this exercise in Fig.~\ref{supfig2}.  As this figure demonstrates, the aHydro freeze-out prescription does the best in reproducing the behavior of all moments.  One sees a modest improvement in the agreement of vHydro when using the LPM-improved prescription, however, it still shows poor agreement at early times where its predictions become negative and unphysical.  This provides some motivation to implement aHydro-type freeze-out for the phenomenological analysis of small collisions systems at RHIC and LHC.

\section{Conclusions}

\vspace{3mm}

In this proceedings contribution, I have presented evidence for the existence of a non-equilibrium attractor in both RTA and EKT QCD kinetic theories.  I demonstrated that in both theories on can identify an attractor in {\em all} moments, that can be extended to early times.  In both cases, the non-equilibrium kinetic-theory attractor smoothly connects to the late-time hydrodynamic attractor.  In addition to this fundamental finding, I also discussed how one can use the determined EKT QCD attractor to test different hydrodynamic freeze-out prescriptions.  This addresses an important open question in the phenomenological analysis of nuclear collisions which is to determine the best prescription for converting hydrodynamical fields into particle distributions. Currently, the quadratic ansatz (i) is widely used. This ansat assumes linear deviations from thermal equilibrium, which is in stark contrast to the far-from-equilibrium conditions in which fluid-dynamical modeling is practiced in current phenomenological applications, in particular in modeling of small systems (see e.g. Refs.~\cite{Bozek:2013uha,Shen:2016zpp,Alqahtani:2016rth,Weller:2017tsr,Mantysaari:2017cni,Strickland:2018exs}).  To address whether these linearized procedures remain quantitatively predictive far from equilibrium, in Ref.~\cite{Almaalol:2020rnu}  we confronted them with far-from-equilibrium simulations of QCD effective kinetic theory.  The results of Ref.~\cite{Almaalol:2020rnu}  show that the aHydro freeze-out ansatz performed better in reconstructing moments of the distribution function compared to linearized ansatze in far-from-equilibrium systems.  Future directions include inclusion of quarks and implementation of aHydro freeze-out into widely used viscous hydrodynamics codes.

\section*{Acknowlegements}

\vspace{1mm}

I thank my collaborators D. Almaalol, A. Kurkela, and U. Tantary.  The work reported here was supported by the U.S. Department of Energy, Office of Science, Office of Nuclear Physics Award No.~DE-SC0013470. 

\section*{References}

\vspace{1mm}

\bibliographystyle{iopart-num}
\bibliography{strickland}

\end{document}